\documentclass[12pt]{article}
\usepackage{epsfig}

\newcommand{\be}{\begin{eqnarray}}
\newcommand{\ee}{\end{eqnarray}}

\begin{document}

\bibliographystyle{unsrt}
\footskip 1.0cm

\thispagestyle{empty}
\begin{flushright}
BCCUNY-HEP-09-05
\end{flushright}
\vspace{0.5in}

\begin{center}{\Large \bf {From RHIC to EIC: Nuclear Structure Functions}}\\

\vspace{1in}
{\large  Jamal Jalilian-Marian}\\

\vspace{.2in}
{\it Department of Natural Sciences, Baruch College, New York, NY 10010\\}
\vspace{0.1in}

and 

\vspace{0.1in}
{\it CUNY Graduate Center, 365 Fifth Ave., New York, NY 10016\\ }

\end{center}

\vspace*{20mm}

\begin{abstract}

\noindent We study the nuclear structure function $F_2^A$ and its logarithmic derivative in the high energy limit (small $x$ region) using the Color Glass Condensate formalism. In this limit the structure function $F_2$ depends on the quark anti-quark dipole-target scattering cross section $N_F (x_{bj}, r_t, b_t)$. The same dipole cross section appears in single hadron and hadron-photon production cross sections in the forward rapidity region in deuteron (proton)-nucleus collisions at high energy, i.e. at RHIC and LHC. We use a parameterization of the dipole cross section, which has successfully been used to describe the deuteron-gold data at RHIC, to compute the nuclear structure function $F_2^A$ and its log $Q^2$ derivative (which is related to gluon distribution function in the double log limit). We provide a quantitative estimate of the nuclear shadowing of $F_2^A$ and the gluon distribution function in the kinematic region relevant to a future Electron-Ion Collider.

\end{abstract}
\newpage

\section{Introduction}
The Color Glass Condensate (CGC) formalism \cite{cgc} has been quite successful in describing aspects of particle production in high energy hadronic/nuclear collisions. It has been used to describe data from lepton-proton (nucleus) Deeply Inelastic Scattering (DIS) to hadron multiplicities and transverse momentum spectra in nucleus-nucleus and deuteron-nucleus collisions \cite{mult}. The recently observed suppression of single hadron $p_t$ spectra in the forward rapidity region in deuteron-nucleus collisions at RHIC \cite{rhic} was predicted by the formalism \cite{predict}. Since then, quantitative results which describe the $p_t$ spectrum of single hadron production in dA collisions have become available \cite{kkt2,dhj,boer}. Therefore it is desirable to apply the knowledge gained from dA collisions at RHIC to other processes in order to further clarify/constrain the region of applicability of CGC formalism to high energy processes \cite{reviews}.

In this work we consider nuclear shadowing of structure function $F_2^A$ and its log $Q^2$ derivative using the results of our previous work on pion production in dA collisions. In \cite{dhj,hybrid} a hybrid approach to dA collisions was used to investigate pion production in dA collisions (for an alternative approach which treats both the projectile and target by CGC formalism, see \cite{alt}). The main result of that analysis \cite{dhj} can be summarized roughly as the following: in the mid rapidity region, as one goes from low to high $p_t$, one goes from the saturation region through the scaling region to the pQCD (DGLAP) region. The applicability of CGC in mid rapidity region of RHIC is limited to less than a few $GeV$ in transverse momentum above which one is in the large $x$ region. It is worth mentioning that the qualitative conclusion reached in \cite{boer} is different from those of \cite{dhj} due to a different parameterization of the dipole cross section which enters the particle production cross section. However, the dipole model in \cite{boer} seems to have an unphysical dependence on transverse momentum at high $p_t$. Therefore, we will use the parameterization of the dipole cross section in \cite{dhj} (referred to  as the DHJ model) in this work.   

In the most forward rapidity region at RHIC (and LHC), one is in the (target) small $x$ kinematics ($p_t$ is limited by kinematics) but in the projectile large $x$ region where the quarks in the projectile contribute more than the projectile gluons (except possibly at very low $p_t$ where they are comparable). Therefore any description of the forward rapidity data must include the effects of the projectile quarks scattering on the gluon field of the target nucleus (or proton).

Using the DHJ model of the dipole cross section employed in \cite{dhj}, we calculate  nuclear structure functions and their shadowing. The main ingredient in the structure function $F_2$ is the (quark anti-quark) dipole cross section which also appears in single hadron production cross section in dA collisions. Specifically, we investigate how the structure function $F_2^A$ computed using the DHJ model of the dipole cross section describes the nuclear shadowing data as measured by the NMC collaboration \cite{nmc}. We then make predictions for $F_2^A$ in a broader kinematic region which may be covered by the proposed Electron-Ion Collider (EIC). We then consider shadowing of the nuclear gluon distribution function. In a high gluon density environment the gluon distribution function (defined as the leading twist matrix element of gluon field operators) is not what goes into a physical cross section, rather, it is $n$-point functions of Wilson lines. Therefore, we consider the log $Q^2$ derivative of the $F_2^A$ structure function which in the high $Q^2$ (double log) limit is proportional to the gluon distribution function. We then compare the ratio of these log derivatives, for a nucleus and a proton, with the results available in the literature.

\section{The structure function $F_2$}

In the small $x$ limit, the structure function $F_2$ can be written as a convolution of the probability for splitting of a virtual photon into a quark anti-quark pair with the probability for the scattering of the quark anti-quark pair on the target so that \cite{raju}
\be
F_2 = a\, Q^4 \, \int d r_t \, r_t \, \int dz \, N_F (x, r_t) \, \bigg\{ f_1 (z)\, K_1^2 (\epsilon\, r_t) + 
f_0 (z) \, K_0^2 (\epsilon\, r_t)\bigg\}
\label{eq:f2}
\ee
where $a = {6\, \sigma_0 \over (2\pi)^3} \sum e_f^2$ is an overall constant and $f_1 (z) = z (1-z) [z^2 + (1-z)^2]$, $f_0 (z) = 4\, z^2 (1-z)^2$. The sum is over quark flavors (we are taking three flavors of massless quarks) and we have defined $\epsilon^2 = z (1-z) Q^2$. All the QCD dynamics in the small $x$ limit is encoded in the dipole cross section $N_F (x, r_t)$ which is also the building block of the single hadron production cross section in the forward rapidity region and satisfies the JIMWLK/BK evolution equations.

It would be ideal to solve the JIMWLK/BK equations in order to obtain the form of the dipole cross section $N_F$ and use it in eq. (\ref{eq:f2}) (see \cite{javier} for a recent analysis of the proton structure function using a solution of the BK equation with running coupling). However, the numerical solutions of JIMWLK/BK equations are known to be quite sensitive to the initial condition in the evolution equation unless one is at extremely small $x$. It is also expected to receive large corrections from higher order corrections. Therefore it is more practical to parameterize the dipole cross section in a form which captures the essential dynamics of the saturation physics. This approach has been employed before to investigate structure functions at HERA \cite{golec} and particle production at RHIC \cite{kkt2,boer,navarra}.

In this work we use the parameterization proposed in \cite{dhj} which was fit to the single hadron production data in dA collisions at RHIC at rapidity $y = 3.2$.  It was then used to predict the data at rapidities  $y =0$ and $y = 4$. For a successful description of the mid rapidity data one had to exclude the contribution from $ x > 0.05$. This is consistent with the lessons learned from HERA where CGC motivated models can not fit the data (for proton targets) for $ x > 0.01$. Therefore, we will restrict ourselves to the region where $x < 0.05$ which puts a severe limit on the number of data points available on nuclear shadowing at reasonably large $Q^2$ (we consider $Q^2 > 1\, GeV^2$ only). There was also a need for a $K$ factor in description of the single hadron spectra in dA collisions which was large (in mid rapidity) but $p_t$ independent (see \cite{jjmK} for observables which are independent of this K factor). This is usually attributed to higher order corrections which are reduced in magnitude as one goes to the forward rapidity region as was the case in \cite{dhj}. Here we do not expect a K factor since we are dealing with a fully inclusive quantity. Furthermore, we do not include the data points measured the E665 collaboration since there seems to be a discrepancy between the NMC and E665 shadowing results when comparing structure functions for large nuclei (such as gold or lead) to that of deuteron which may be due to the systematic errors of the E665 measurement \cite{antje}.

\subsection{The dipole cross section and shadowing of $F_2^A$}

In DHJ parameterization \cite{dhj}, the fundamental (quark anti-quark) dipole scattering cross section is given by (an overall factor of $\sigma_0 = 23\, mb$ is factored out in this expression but included in the numerical results)
\be  
\label{eq:NF_param}
N_F(x, r_t) = 1-\exp\left[ - \frac{1}{4} [r_t^2 Q_s^2(x)]^
{\gamma(x,r_t)}\right]
\ee
where the anomalous dimension $\gamma$ is given by (for details see \cite{dhj})
\be
\gamma(r_t,y) &=& \gamma_s + \Delta\gamma(r,y) \nonumber\\
\Delta\gamma &=& (1-\gamma_s)\,\frac{\log (1/r_t^2\,Q_s^2)}{\lambda\, y
+\log(1/r_t^2\,Q_s^2) + d\sqrt{y}}~.  \label{eq:gam_new} 
\ee 
where $y \equiv \log 1/x$ and $d = 1.2$ (see \cite{dhj} for details of the parameterization). Due to the anomalous dimension $\gamma$, there is a strong "leading twist" shadowing of the target gluons encoded in this formalism \cite{dilep}. Using this, we compute the minimum bias structure functions for a proton and a nucleus (taken to be gold) in the kinematic region covered by the NMC experiment while keeping $x < 0.05$ and $Q^2 > 1\, GeV^2$. 

\begin{figure}[hbtp]
  \begin{center}
   \includegraphics[width=4.5in]{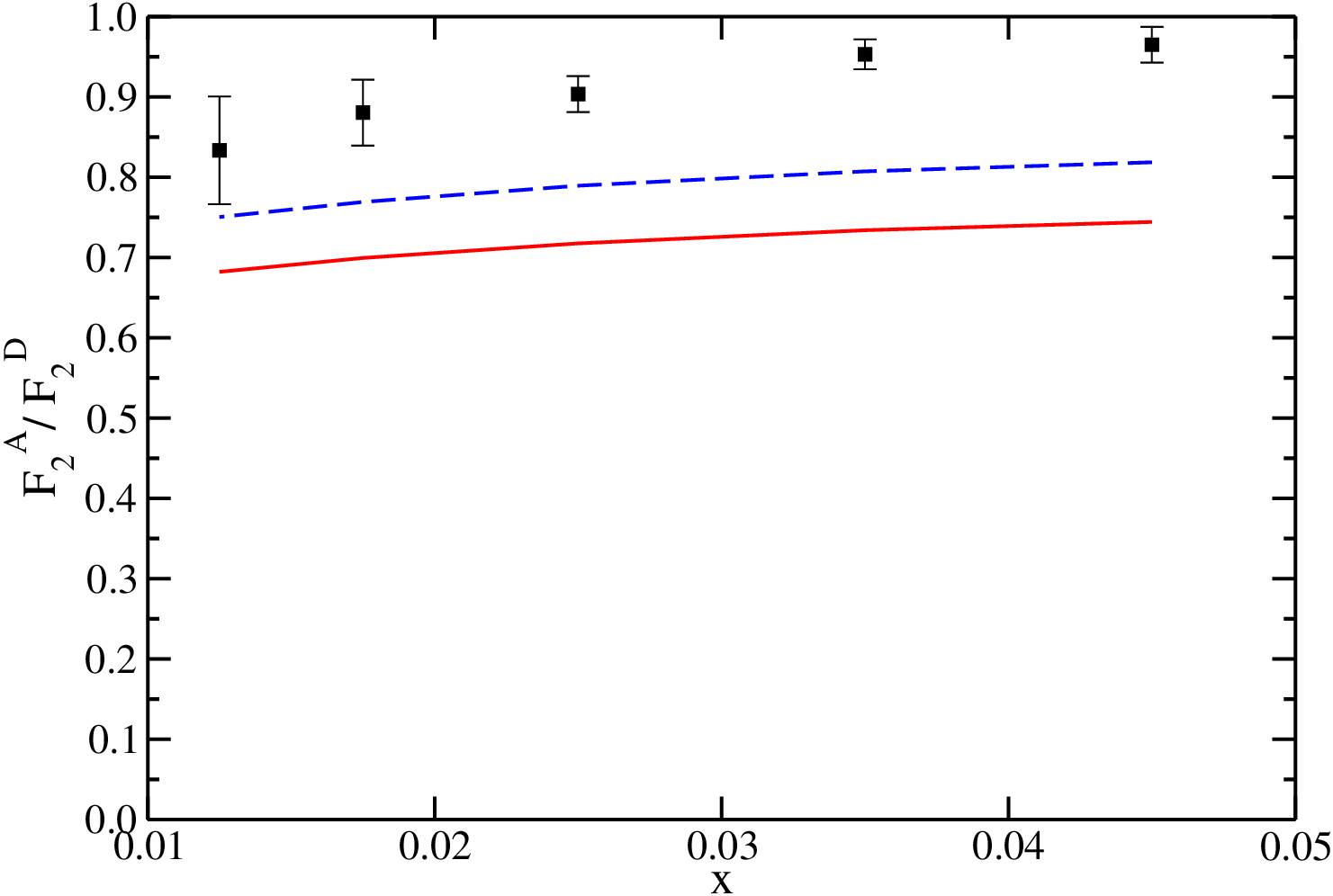}
   \caption{Shadowing of structure function $F_2^A$ compared to NMC measurements.}
   \label{fig:R_F2_mod}
   \end{center}
   \end{figure}

Our result for nuclear shadowing is shown in Fig. (\ref{fig:R_F2_mod}). The lower (solid) line is the ratio of $F_2^A/F_2^p$ where both structure functions for proton and gold are calculated using this formalism. The upper (dashed) line is the ratio $F_2^A/F_2^d$ where we have included an overall $10\%$ uncertainty to account for the effect of deuteron shadowing (about $2-3\%$ in this kinematics \cite{e665}), the fact that the experimental data is for lead rather than gold and the fact that our calculation of the proton structure function overestimates the HERA data by about $10\%$ \cite{navarra}. All these effects would make the calculated shadowing ratio increase. The fact that we are over estimating the proton structure function is not unreasonable since the DHJ parameterization was developed and tested for nuclear rather than proton targets. We emphasize that there is no new parameters introduced for the computation of the structure function. Alternatively, we could have divided the nuclear structure function $F_2^A$ calculated using this formalism by the proton structure function measured at HERA. Since we will consider the shadowing of the "gluon distribution function" defined as the log $Q^2$ derivative of $F_2$, we have decided to divide by the calculated value of $F_2^p$ rather than the measured value and thus have included the upper line as a more realistic estimate of the shadowing effect. As is clear, the difference between the data points is largest (of order of $15\%$) at the highest $x$ considered ($x = 0.05$) and gets much smaller (a few percent) toward the lowest $x$ in the figure ($x = 0.01$).

\begin{figure}[hbtp]
   \begin{center}
   \includegraphics[width=4.5in]{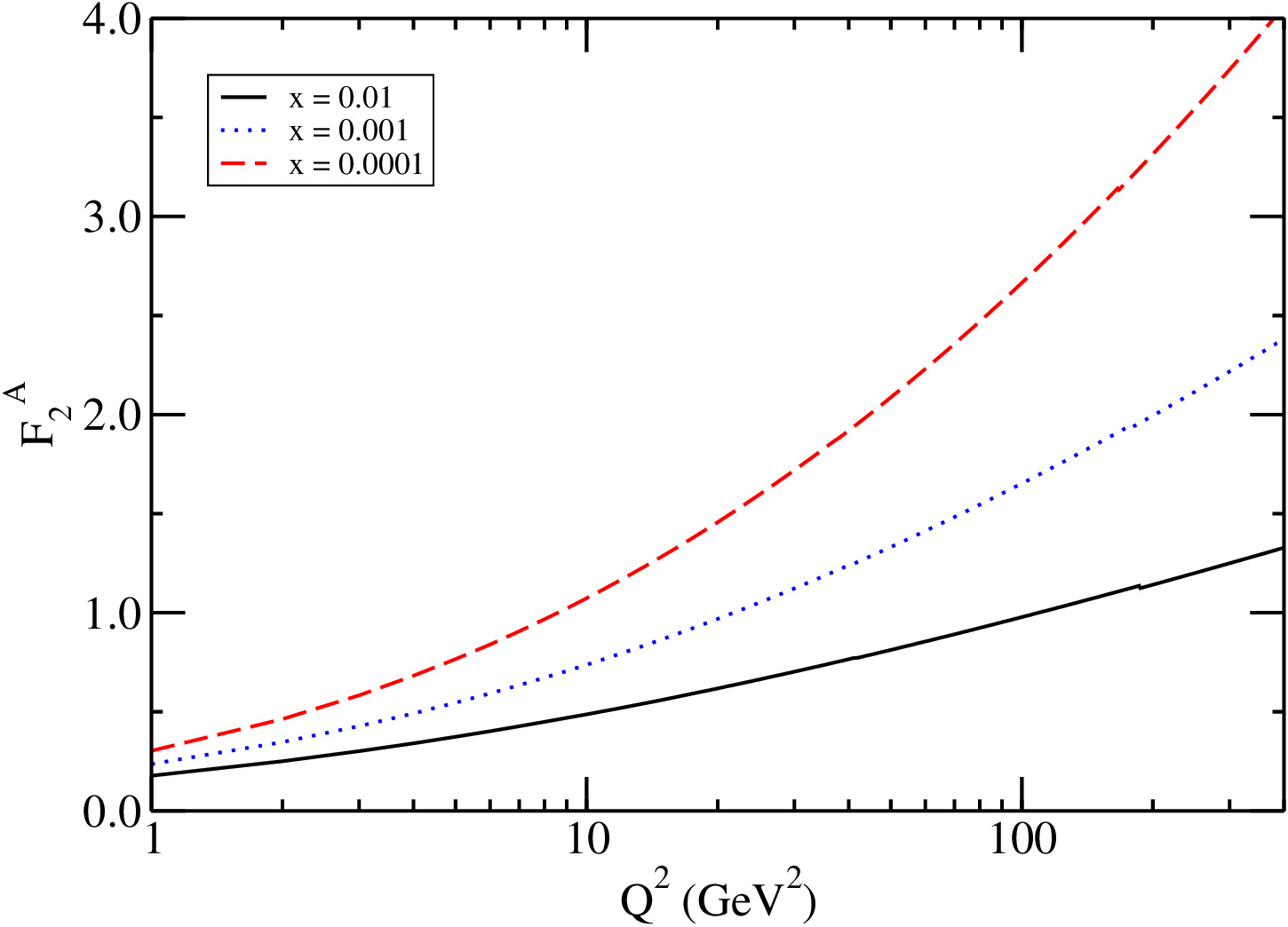}
   \caption{Nuclear structure function $F_2$.}
   \label{fig:F2A_vs_Q2}
   \end{center}
   \end{figure}

In Figure (\ref{fig:F2A_vs_Q2}) we show the nuclear structure function $F_2^A$ as a function of $Q^2$ at different values of $x$. We show our results for values of $Q^2$ up to $400\, GeV^2$ which was an estimate for the absolute upper limit of the scaling region at HERA \cite{scaling}. The analysis in \cite{dhj} shows that the scaling region may actually be smaller. Furthermore, the calculation of $F_2^A$ is more reliable in our case since we don't have to consider the proton structure function which our calculation overestimates by about $10\%$.

\vspace{0.3in}

\begin{figure}[hbpt]
   \begin{center}
   \includegraphics[width=4.5in]{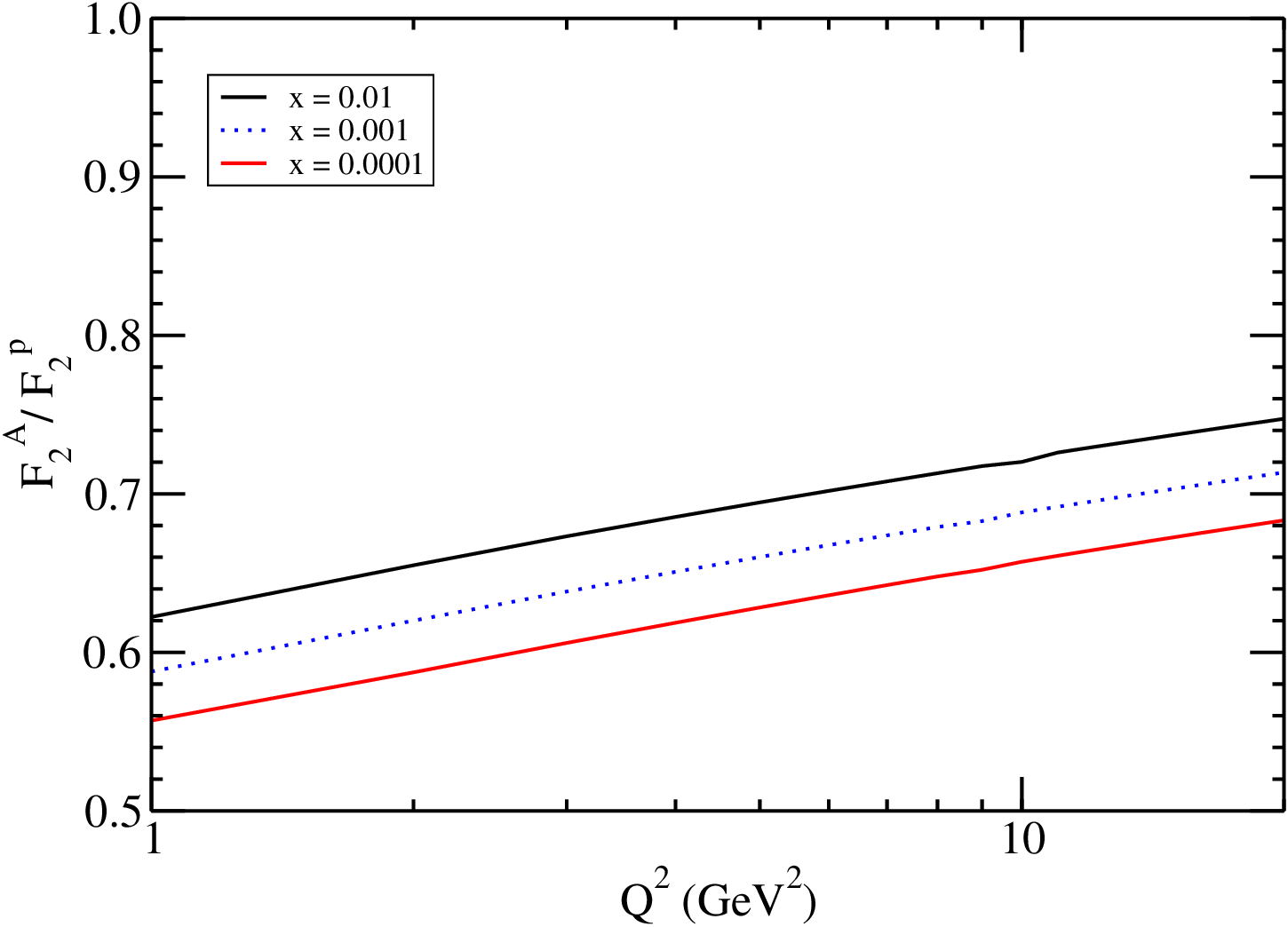}
   \caption{Nuclear shadowing of the structure function $F_2^A$.}
   \label{fig:R_F2A_F2p_vs_logQ2_shortest}
   \end{center}
   \end{figure}

\begin{figure}[hbtp]
   \begin{center}
   \includegraphics[width=4.5in]{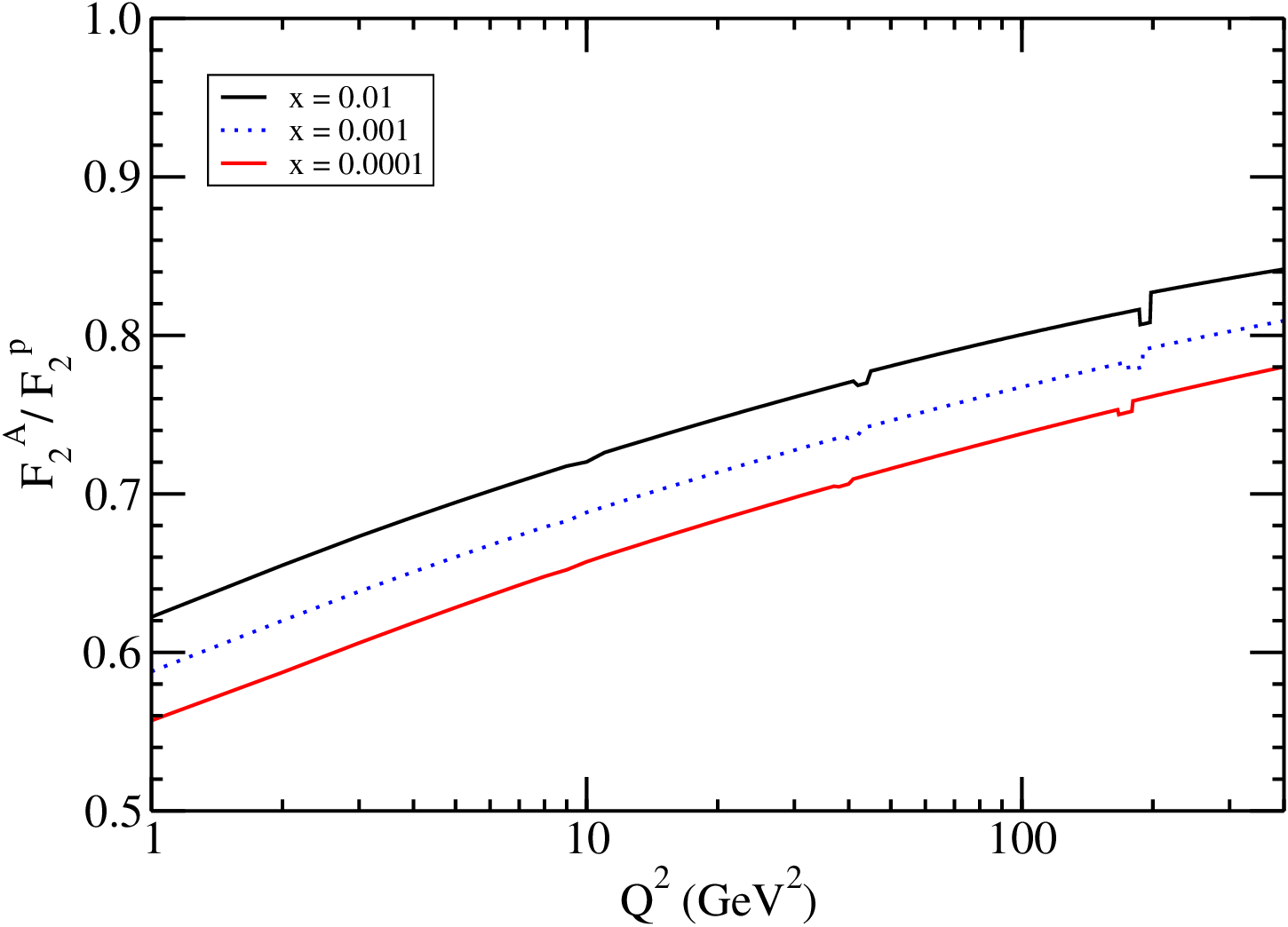}
   \caption{Same as in Fig. (\ref{fig:R_F2A_F2p_vs_logQ2_shortest}) but in a larger kinematic region.}
   \label{fig:R_F2A_F2p_vs_logQ2}
   \end{center}
   \end{figure}

In Figures (\ref{fig:R_F2A_F2p_vs_logQ2_shortest}) and  (\ref{fig:R_F2A_F2p_vs_logQ2}) we show the shadowing effect for the structure function at different kinematics. In both figures we have plotted the ratio of $F_2$ for minimum bias gold nucleus and that of a proton and have ignored the shadowing of deuteron structure function at small $x$ (which may eventually be about $10\%$) and the fact that our calculation overestimates the proton structure function at higher $Q^2$ by about $10\%$. Therefore, the lines in both figures should be understood to be a lower limit which could be pushed up by $10- 15\%$ in a more detailed study. 

It is worth noting that one could fit the $Q^2$ dependence of the lines shown in Fig. (\ref{fig:R_F2A_F2p_vs_logQ2_shortest}) by a function of the form $ c_0 + c_1 Log \, Q^2/Q_s^2 $ with constants $c_0, c_1$ being the same (with $2\%$ accuracy) for all three lines shown (the deviation is largest at smallest $Q^2$ which may be a sign of reaching the saturation region). Whether the saturation scale appearing in the log is that of a proton or nucleus can not be determined from the fit since the constant $c_1$ is much smaller than $c_0$. It is tempting to take this as a sign that the shadowing in this kinematics is "leading twist", consistent with the estimates of the scaling region where the dynamics is that of BFKL with saturation boundary (we note that the saturation scale of a minimum bias gold nucleus is $~ 1.6\, GeV $ at $x = 3\times 10^{-4}$ and $~1\, GeV$ for a proton). On the other hand, in the much larger kinematics shown in Fig. (\ref{fig:R_F2A_F2p_vs_logQ2}) this functional form does not provide a very accurate fit which may be a sign that we are outside the scaling region. 

It would be very interesting to make analytic estimates for the extent of the scaling region in this case, however, this is a bit complicated due to the $z$ integration in eq. (\ref{eq:f2}). In principle, the value of photon virtuality $Q^2$ determines whether one is in the saturation region $Q \ll Q_s$, in the scaling region $Q_s \ll Q \ll Q_s^2/\Lambda_{QCD}$ or in the pQCD region where $Q_s^2/\Lambda_{QCD} \ll Q$.
The the integration over the dipole size $r_t$ can be broken up into $4$ distinct regions
\be
\int_0^{\infty} r_t = \int_0^{\Lambda/Q_s^2} \, + \, \int_{\Lambda/Q_s^2}^{1/Q} \, + \, \int_{1/Q}^{1/Q_s} \, + \, \int_{1/Q_s}^{\infty} 
\label{eq:regions}
\ee
The first region is that of pQCD where $r_t\, Q \ll 1$ and $r_t\, Q_s \ll 1$. Then $N_F \sim r_t^2\, Q_s^2$ and $K_1^2 \sim 1/\epsilon^2\, r_t^2$ and $K_0^2$ is subleading. In this case the $r_t$ and $z$ integrations can be performed analytically and are finite. A similar thing happens in the saturation region (the last piece in eq. (\ref{eq:regions})). In this case, $r_t\, Q \gg 1$ and $r_t\, Q_s \gg 1$ and one can approximate $N_F \sim 1$ and $K_0^2 \sim K_1^2 \sim (1/\epsilon\, r_t) \, exp (-2 \epsilon\, r_t\, Q)$. However, in the other two regions, the $z$ integration diverges as $z\rightarrow 0, 1$ and is not under control (having massive quarks would regulate this divergence). One can still perform the integrals analytically (for fixed $\gamma$) but its general form is not very illuminating.

\subsection{Shadowing of gluons}

In this section we consider nuclear shadowing of the gluons. In pQCD and at the leading twist level, the gluon distribution function $xG$ is proportional to the log $Q^2$ derivative of $F_2$ structure function,
\be
xG (x, Q^2) \sim {d\over d \, Log\, Q^2} F_2 (x, Q^2)
\label{eq:xg}
\ee
This relation can be used to extract the gluon distribution function experimentally in DIS. The Color Glass Condensate formalism, however, is not based on a twist expansion and the relevant degrees of freedom (which appear in physical cross sections) are not gluons but Wilson lines, path ordered exponentials of gluon fields which take multiple scattering on the target into account. In a typical cross section computed in the CGC formalism, one encounters $n$-point functions of Wilson lines (fundamental or adjoint), where in case of the structure function $F_2$, only the two point function of (fundamental) Wilson lines appears. Therefore, in order to provide a comparison with results obtained from other approaches based on leading twist collinear factorization (where shadowing is put in by hand in the initial condition), we consider the log $Q^2$ derivative of the $F_2$ structure function and take that to be a measure of shadowing of gluons in the standard language.

\begin{figure}[hbtp]
   \begin{center}
   \includegraphics[width=4.5in]{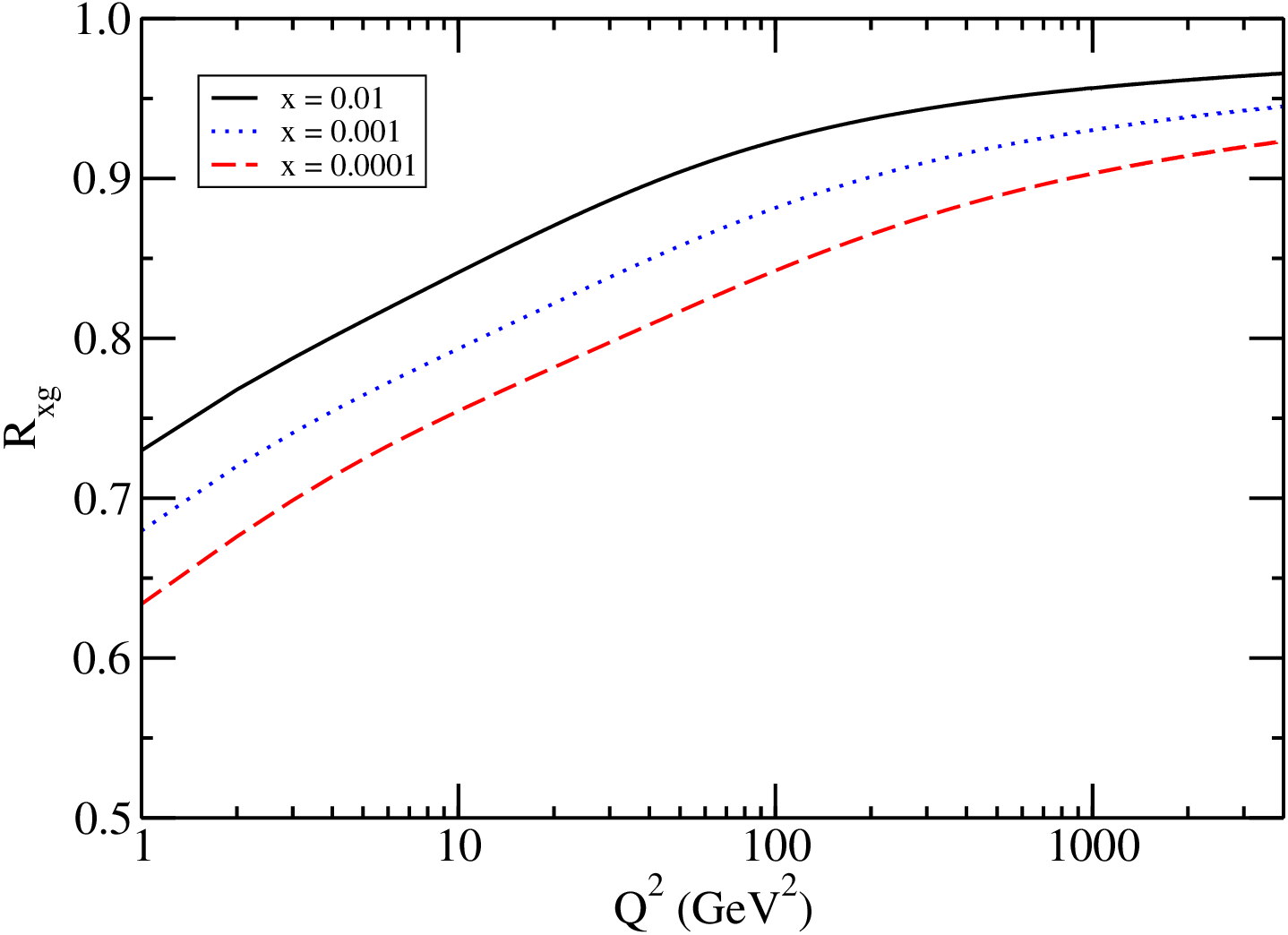}
   \caption{Shadowing of the gluon distribution function as defined in eq. (\ref{eq:R_xg}).}
   \label{fig:R_xg_vs_logQ2}
   \end{center}
   \end{figure}

\begin{figure}[hbtp]
   \begin{center}
   \includegraphics[width=4.5in]{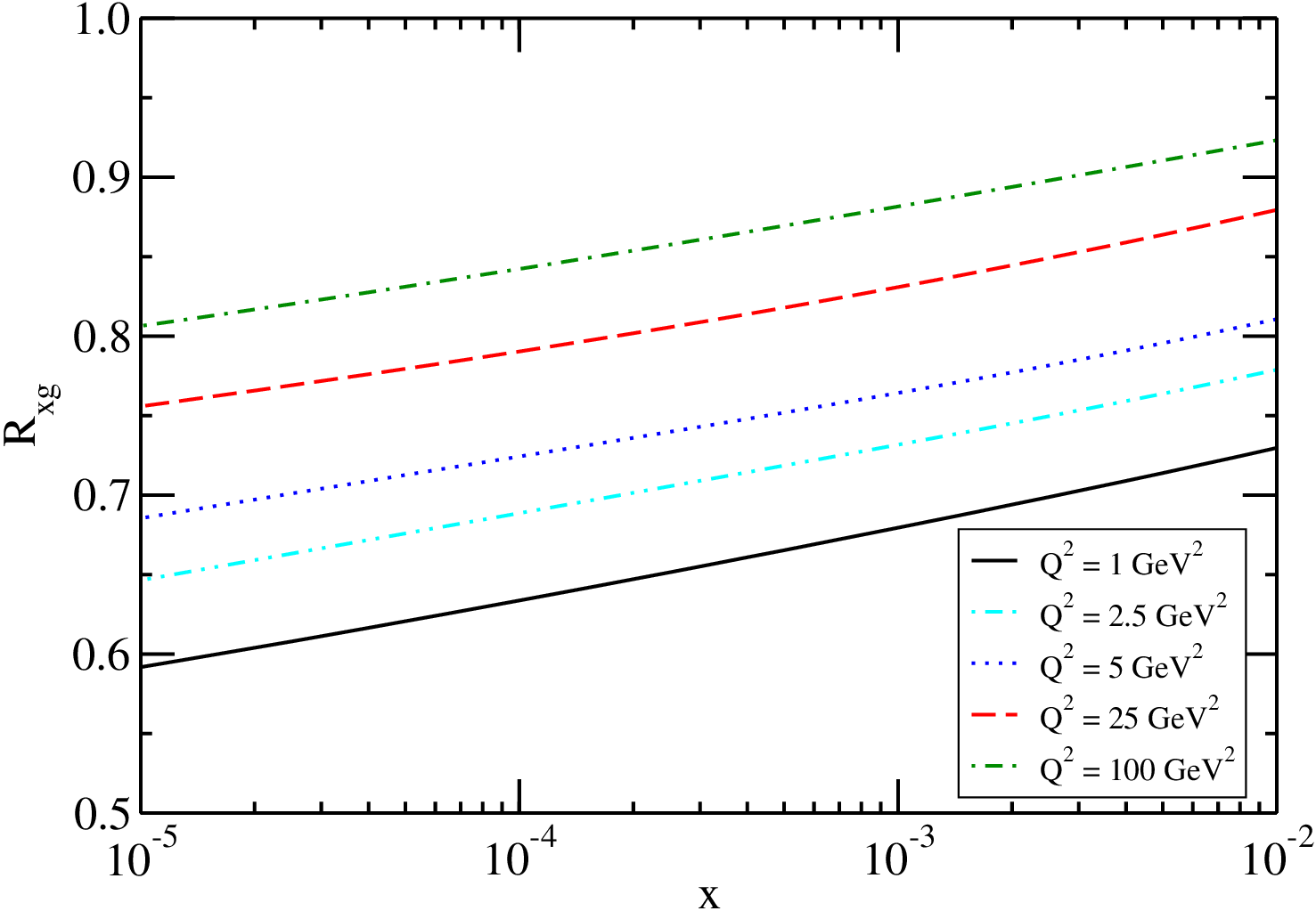}
   \caption{Same as in Fig. (\ref{fig:R_xg_vs_logQ2}) but vs. $x$.}
   \label{fig:R_xg_vs_x}
   \end{center}
   \end{figure}

Differentiating the structure function $F_2$ as given in eq. (\ref{eq:f2}) with respect to log $Q^2$, we get two terms; one is coming from differentiating the overall $Q^4$ term and the second contribution is coming from the $Q^2$ dependence of the Bessel functions. We get
\be
{d\, F_2 \over d\, log\, Q^2} &=& a\, Q^4 \, \int d r_t \, r_t \int d z \, N_F (x, r_t) \bigg\{
2 f_0 (z) K_0^2 (\epsilon\, r_t) - \nonumber\\
&& \epsilon\, r_t\, [f_0 (z) + f_1 (z)]\, K_0 (\epsilon\, r_t) K_1 (\epsilon\, r_t)  + f_1 (z) \, K_1^2 (\epsilon\, r_t)\bigg\}.
\label{eq:f2der}
\ee 
Using the expression for $N_F$ given by (\ref{eq:NF_param}) we can now calculate the log derivative of the structure function. In Figs. (\ref{fig:R_xg_vs_logQ2}, \ref{fig:R_xg_vs_x}) we show $R_{xg}$ defined as 
\be
R_{xg} \equiv {d\, F_2^A \over d\, log\, Q^2}/{d\, F_2^p \over d\, log\, Q^2}.
\label{eq:R_xg}
\ee 

Here we have ignored shadowing of gluons in a deuteron. We are also using our formula to calculate this log derivative for a proton target rather than using $xg$ from available parameterizations such as CTEQ or MRST for self consistency. Because of this we may underestimate the amount of shadowing by $\sim 10 - 15\%$ in the small $x$ but large $Q^2$ region. This would mean, for example in Fig. (\ref{fig:R_xg_vs_logQ2}) that this ratio is very close to $1$ at the highest $Q^2$ shown. At the largest $x$ considered ($x = 0.01$), the amount of gluon shadowing is about $10 - 15\%$ at $Q^2 \sim 10\, GeV^2$ and $25\%$ at the $Q^2 = 1\, GeV^2$. It goes away by $Q^2$ of several hundred $GeV^2$.  One should keep in mind that the shadowing results shown are most likely an overestimate (the real ratios are a bit larger) since first, we are ignoring deuteron shadowing and second, we are overestimating the gluon distribution in a proton by possibly $10\%$. At smaller $x$ and $Q^2 = 1\, GeV^2$ gluon shadowing is about $30\%$ and again goes away at higher $Q^2$. The $Q^2$ dependence of shadowing at fixed $x$ is shown in Fig. (\ref{fig:R_xg_vs_x}).

\section{Summary}

We have used the parameterization of the quark anti-quark dipole successfully employed in \cite{dhj} to investigate the predictions of the model for shadowing of minimum bias nuclear structure functions. We have compared our results to the available data from NMC collaboration in the smallest $x$ region accessed in that fixed target experiment while keeping the values of $Q^2$ larger than $1\, GeV^2$. Our results are within a few percent of the data at the smallest $x$ bin ($\sim 0.0125$) and are off by about $15\%$ at the highest $x$ considered ($\sim 0.045$). Defining gluon distribution function as the log $Q^2$ derivative of the $F_2$ structure function, we calculated the shadowing of gluons and have made predictions for its $x$ and $Q^2$ dependence which could be tested in a future lepton-nucleus collider experiment.

It is interesting to compare our results with those from other approaches. Typically there are two distinct parton-based approaches (see also \cite{mark} for an alternative approach) to particle production in high energy collisions and nuclear shadowing of structure functions. First is based on the leading twist collinear factorization and DGLAP evolution of distribution functions. In this approaches one has to put in shadowing by  hand as a modification of the distribution functions at some scale $Q_0^2$ which are then evolved according to DGLAP equations \cite{hkn,eks98,eks08}. Attempts to simultaneously fit the RHIC data on forward rapidity dA collisions and the nuclear structure functions seem to lead to an almost $100\%$ shadowing of gluons in a gold or lead nucleus at small $x$ which seems quite unnatural \cite{eks08} (this ratio seems to be higher in the most recent version EPS09 \cite{eks08} which uses a NLO analysis). Comparing our results for shadowing of gluon distribution function as defined in eq. (\ref{eq:f2der}) to those in \cite{eks98,eks08}, our shadowing ratio is much closer to the rsults of \cite{eks98} than \cite{eks08}. For example, we predict $R_{xg} \sim 0.65$ (for minimum bias gold nucleus) at $x = 10^{-4}$ and $Q^2 \sim 1.7\, GeV^2$ while this ratio in \cite{eks98} (LO) is $\sim 0.5$, but $\sim 0$ in \cite{eks08} and $\sim 0.5$ in the most recent version in \cite{eks08} (NLO). The difference in the amount of gluon shadowing between \cite{eks98} and \cite{eks08} (LO) is quite interesting. The drastic shadowing of gluons in \cite{eks08} (there are almost no gluons in the nucleus) is caused by insisting on describing the RHIC forward rapidity data using collinear factorization approach where nuclear shadowing is put in the initial gluon distribution by hand. Since having almost no gluons in the nucleus seems a bit unnatural, this may indicate that leading twist collinear factorization approach to forward rapidity data at RHIC is not valid. The amount of gluon shadowing in the NLO version is much less than in the LO version but it is not clear whether the NLO version also reproduced the forward rapidity RHIC data. There has also been attempts to include the effect of the higher twist corrections to the DGLAP evolution equations \cite{glr} but these effects would break collinear factorization and make the approach practically useless for hadronic/nuclear collisions.

The second approach is based on the concept of gluon saturation and here also one can distinguish between two different approximations to the dipole cross section (see also \cite{f2scaling} which studied shadowing using scaling ideas). One is motivated by the original approach of Golec-Biernat and Wusthoff which is a Glauber like multiple scattering approach to shadowing but using the parton language. The shadowing generated here is a "higher twist" (suppressed by powers of $Q^2$) effect. One such work is \cite{golec} (last paper) which also investigates the impact parameter dependence of shadowing, but does not include the effects of BFKL anomalous dimension $\gamma$ and in this sense does not have "leading twist" shadowing. A second approach within the saturation picture includes this anomalous dimension but using a different model for the dipole cross section \cite{navarra}. It is also used to fit the RHIC data in the forward rapidity data but to the best of our knowledge it has not been compared against the NMC data on shadowing of $F_2$ (it has been compared with the E665 data however) and shadowing of gluons is not considered.

The fact that our parameter free calculation of the $F_2$ structure function is so close to the experimental values is another indication that the observed suppression of the single hadron spectra in the forward rapidity dA collisions is due to gluon saturation dynamics. Nevertheless there are some interesting questions which have not been addressed here and need further investigation. In this context, the first and most important is probably a study of the impact parameter dependence of structure functions including BFKL dynamics. The DHJ parameterization of the dipole cross section was proposed for minimum bias dA collisions. It would be interesting to do a similar analysis for the centrality dependent hadron spectra in dA collisions and apply the knowledge to predict the centrality dependence of the structure functions which could be measured in a future facility. This would help clarify the dynamics of shadowing, whether is it power or logarithmically suppressed. Since the magnitude of saturation scale is much larger for most central collisions, then the saturation region for central collisions extends to higher values of $Q^2$ as compared with minimum bias collisions where the value of $Q_s^2$ is not that large. This work is in progress and will be reported elsewhere \cite{adjjm_bt}.

One could also differentiate between different approaches (CGC vs. collinear factorization inspired) to shadowing by considering two particle correlations \cite{twopart}. In the CGC approach where one describes both the projectile and target using CGC formalism, one expects a de-correlation of the two particles as the rapidity separation between the two observed particles gets larger. This is due to the small $x$ evolution of the gluon ladder between the produced hadrons. In the standard collinear factorized form of particle production where shadowing is put in by hand, there should be no de-correlation. Therefore, a detailed study of $p_t$, impact parameter  and rapidity dependence of two particle production would go a long way toward clarifying the dynamics of shadowing of nuclear structure functions.

\vspace{0.2in}
\leftline{\bf Acknowledgments} 

\noindent We thank A. Dumitru for discussions related to this work and V. Guzey and M. Strikman for correspondence regarding deuteron shadowing.

\vspace{0.2in}
\leftline{\bf References}

\renewenvironment{thebibliography}[1]
        {\begin{list}{[$\,$\arabic{enumi}$\,$]}  
        {\usecounter{enumi}\setlength{\parsep}{0pt}
         \setlength{\itemsep}{0pt}  \renewcommand{\baselinestretch}{1.2}
         \settowidth
        {\labelwidth}{#1 ~ ~}\sloppy}}{\end{list}}

\end{document}